\documentclass[showpacs,prb,twocolumn]{revtex4}
\usepackage{amsfonts}
\usepackage{amssymb}
\usepackage{amsmath}
\usepackage{graphicx}


\begin{document}

\title{Competition between charge and spin order in the $t-U-V$ extended Hubbard model on the triangular lattice.}
\author{B.~Davoudi$^{1,2}$, S.~R.~Hassan $^{1}$ and A.-M.S.~Tremblay$^{1}$}
\begin{abstract}
Several new classes of compounds can be modeled in first approximation by electrons on the triangular lattice that interact through on-site repulsion $U$ as well as nearest-neighbor repulsion $V$. This extended Hubbard model on a triangular
lattice has been studied mostly in the strong coupling limit for only a few types of instabilities. Using the extended 
two-particle self consistent approach (ETPSC), that is valid at weak to intermediate coupling, we present an unbiased 
study of the density and interaction dependent crossover diagram for spin and charge density wave instabilities of the 
normal state at arbitrary wave vector. When $U$ dominates over $V$ and electron filling is large, instabilities are 
chiefly in the spin sector and are controlled mostly by Fermi surface properties. Increasing $V$ eventually leads to 
charge instabilities. In the latter case, it is mostly the wave vector dependence of the vertex that determines the wave 
vector of the instability rather than Fermi surface properties. At small filling, non-trivial instabilities appear only 
beyond the weak coupling limit. There again, charge density wave instabilities are favored over a wide range of dopings by
large $V$ at wave vectors corresponding to $\sqrt(3) \times \sqrt(3)$ superlattice in real space. Commensurate fillings 
do not play a special role for this instability. Increasing $U$ leads to competition with ferromagnetism. At negative 
values of $U$ or $V$, neglecting superconducting fluctuations, one finds that charge instabilities are favored. 
In general, the crossover diagram presents a rich variety of instabilities. We also show that thermal charge-density 
wave fluctuations in the renormalized classical regime can open a pseudogap in the single-particle spectral weight, 
just as spin or superconducting fluctuations.

\end{abstract}

\affiliation{$^{1}$D\'{e}partment de Physique and RQMP, Universit\'{e} de Sherbrooke,
Sherbrooke, Qu\'{e}bec, Canada J1K 2R1\\
$^{2}$Institute for Studies in Theoretical Physics and Mathematics, Tehran
19395-5531, Iran}
\maketitle

\section{Introduction}
One of the outstanding problems in quantum many-body physics is to understand quasi-two dimensional systems where both electron-electron interaction and geometric frustration are important~\cite{Ong}. The triangular lattice is the prime example where the geometry frustrates near-neighbor anti-alignment of the spins that naturally tends to occur in the presence of short-range electron-electron interaction. Studying models of interacting electrons on such a lattice is thus certainly of fundamental interest, but it is also strongly motivated by the discovery of new materials. Prime examples of these materials are organic bis-(ethylenedithio) (BEDT) Cu$_2$(CN)$_3$ layered compounds~\cite{Shimizu:2003}, triangular lattice antiferromagnets of the CuCrO$_2$ family~\cite{Seki:2008}, and transition-metal oxide materials like ${\rm Na}_x {\rm CoO}_2$ and ${\rm Na}_{1-x} {\rm TiO}_2$. The layered cobaltates have drawn much attention because of their unconventional properties. Sodium cobaltate shows an unusually strong thermopower\cite{Terasaki} at doping $x\approx 2/3$ that can be suppressed drastically by applying an in-plan magnetic field\cite{Wang}. The observation of Curie-Weiss behavior in the magnetic susceptibility while resistivity displays metallic behavior is another puzzle\cite{Ray}. The system also becomes superconductor when it is diluted by water\cite{Schaak,Chou,Jin}. Various types of charge- and spin orders also have been found in the system for various dopings\cite{Foo, Tokura, Mackenzie}.

${\rm Na}_{x}{\rm CoO}_{2}$ consist of two-dimensional ${\rm CoO}_{2}$ layers separated by insulating ${\rm Na}^{2+}$ layers. The ${\rm CoO}_{2}$ layers have Co atoms at the center of oxygen octahedra forming a 2D triangular lattice. The band structure calculation performed by Singh\cite{Singh}, revealed details of splitting of the $3d^5$ bands in Co atoms.  With help of this calculation and also of NMR experiments\cite{Ray}, one can find a rough estimate of hopping and exchange constants that would enter a two-dimensional Hubbard or $t-J$ model for this system. However, the modeling is complicated by the fact that band structure calculations lead to hole pockets that are not observed experimentally, a question that is still debated by several groups using, for example, the Gutzwiller approximation~\cite{Zhou:2005}, the local density approximation plus Hubbard~\cite{Shorikov:2007} $U$ and dynamical-mean field theory~\cite{Liebsch:2007,Marianetti:2006,Aryanpour:2006}. In addition, the effect of long-range Coulomb interaction from the sodium leads to modifications to the simplest Hubbard Hamiltonian for the cobaltates\cite{Roger:2007,Julien:2007}.

In this paper, we do not address the question of detailed modeling of the cobaltates or of other triangular lattice systems. Instead, we note that since several types of spin and charge density waves are observed in these materials, it is quite likely that first-neighbor repulsion $V$, and not only on-site repulsion $U$, must be taken into account. $U$ by itself tends to favor spin-density waves. We thus just focus on the simplest extended $t-U-V$ one-band model Hubbard model on the triangular lattice and ask a few general questions: What types of phases are typical in different doping ranges, what type of interaction favors them, and should one expect pseudogap effects.

Previous theoretical and numerical works have obtained phase diagrams for the triangular lattice in the presence of competing interactions. There are, for example, variational Monte-Carlo calculations~\cite{Watanabe,Motrunich1} for the extended Hubbard model. That work focused mostly on the presence of the Charge density wave (CDW) at filling $n=2/3$ and RVB superconductivity at $n=1/3$. Slave boson methods were used for the $t-V$ and $t-J$ models~\cite{Motrunich2,Baskaran} to study CDW, ferromagnetism and also RVB superconductivity. Series expansion methods and cluster mean field theory~\cite{Zheng} have also investigated CDW, N\'eel order, ferromagnetic order, dimer order and phase separation in a $t-J-V$ model. We will comment further on some of these calculations in the context of our own results.

The results of this paper are obtained with the recently developed Extended two-particle self-consistent approach~\cite{Bahman1,Bahman2}(ETPSC) that is valid from weak to intermediate coupling. This method has been benchmarked against Quantum Monte-Carlo (QMC) simulations (QMC) for the extended Hubbard model on a square lattice. The approach satisfies conservation laws and the Mermin-Wagner theorem stating that no continuous symmetry can be broken at finite temperature in two dimensions. More traditional methods, such as the Random phase approximation, do not satisfy this requirement. With ETPSC, quantum renormalization of interactions (Kanamori-Br\"{u}uckner screening) is taken into account. Instability towards zero-temperature long-range order is signaled at finite temperature by crossover to the renormalized-classical regime where the correlation length grows exponentially. The wave-vector of the instability is determined self-consistently by the approach and all wave vectors are in principle allowed. No a priori selection is necessary.

Within ETPSC we can also compute the self-energy and other related quantities, such as the spectral weight that is measured in photoemission experiments~\cite{AM2}. For the Hubbard model, it has been shown with the Two-particle self-consistent approach (TPSC) that a pseudogap can appear as precursor induced by either antiferromagnetic~\cite{AM2} or superconducting fluctuations~\cite{AM2,Kyung:2001}. The former~\cite{Dare:2004} has been observed experimentally in electron-doped high-temperature superconductors~\cite{Motoyama:2007}. Our results demonstrate that CDW fluctuations can also induce a pseudogap. This is a relevant question experimentally given that CDW induced pseudogaps are observed and sometimes even show similarities with observations in high-temperature superconductors~\cite{Kordyuk:2008}.

In the following we first introduce the model and the ETPSC methodology. We next present our numerical results, discussing various physical effects in terms of the spin and charge structure factors. We display phase diagrams that help understand how microscopic parameters favor various phases. The CDW induced pseudogap and its effect on the Fermi surface are discussed before we present an overview and a conclusion.

\section{Model and method}
\label{sec2}

We write the extended Hubbard Hamiltonian in the following form,%
\begin{widetext}
\begin{equation}
H=-t\sum_{\left\langle \mathbf{ij}\right\rangle \sigma }(c_{\mathbf{i}\sigma
}^{\dagger }c_{\mathbf{j}\sigma }+c_{\mathbf{j}\sigma }^{\dagger }c_{\mathbf{%
i}\sigma })+U\sum_{\mathbf{i}}n_{\mathbf{i}\uparrow }n_{\mathbf{i}\downarrow
}+V\sum_{\left\langle \mathbf{ij}\right\rangle \sigma \sigma ^{\prime }}n_{%
\mathbf{i}\sigma }n_{\mathbf{j}\sigma ^{\prime }}-\mu \sum_{\mathbf{i}}n_{i}
\end{equation}%
\end{widetext}
where $c_{\mathbf{i}\sigma }$ ($c_{\mathbf{i}\sigma }^{\dagger }$) are
annihilation (creation) operators for electrons of spin $\sigma $ at site $i$
of a triangular lattice, $n_{\mathbf{i}\sigma }$ is the density operator, and $t$ is the hopping matrix element. The quantities $U$ and $V$ are the on-site and nearest-neighbor interactions respectively and $\mu $ is the chemical
potential.

For the Hubbard model ($V=0$), TPSC is a very reliable approach up to intermediate coupling limit. The functional derivative method is particularly convenient to obtain the TPSC equations~\cite{Allen:2003}. This is the method that was used to generalize TPSC to the extended Hubbard model \cite{Bahman1, Bahman2}.

The equations that need to be solved are the following. The charge and spin response functions take the form
\begin{equation}
\label{chi_cc}
\chi _{cc}(\mathbf{q},\omega _{n})=\frac{\chi ^{0}(\mathbf{q},\omega _{n})}{
1+\frac{\chi ^{0}(\mathbf{q},\omega _{n})}{2}U_{cc}(\mathbf{q})}
\end{equation}
\begin{equation}
\label{chi_ss}
\chi _{ss}(\mathbf{q},\omega _{n})=\frac{\chi ^{0}(\mathbf{q},\omega _{n})}{
1-\frac{\chi ^{0}(\mathbf{q},\omega _{n})}{2}U_{ss}(\mathbf{q})}
\end{equation}
where
\begin{widetext}
\begin{align}
\label{vertex}
U_{cc}({\mathbf q})&=U\left(g_{\sigma\tilde{\sigma}}(0)+n \frac{\delta g_{\sigma\tilde{\sigma}}(0)}{\delta n}\right) +4V\left(g_{cc}(a)\gamma({\mathbf q})+n\frac{\delta g_s(a)}{\delta n}(3+\gamma({\mathbf q}))\right),\notag\\
U_{ss}({\mathbf q})&=Ug_{\sigma\tilde{\sigma}}(0)- 4V\left(g_{ss}(a)\gamma({\mathbf q})+3n\frac{\delta g_s(a)}{\delta m}\right),
\end{align}
\end{widetext}
are the charge and spin vertex functions and $\chi ^{0}(\mathbf{q},\omega _{n})$ is the free response function (non-interacting susceptibility) given by
\begin{equation}
\label{Chi_0}
\chi ^{0}(\mathbf{q},\omega _{n})=\int_{BZ}\frac{d\mathbf{p}}{\nu }\frac{
f^{0}(\mathbf{p}+\frac{\mathbf{q}}{2})-f^{0}(\mathbf{p}-\frac{\mathbf{q}}{2})
}{i\omega _{n}-\epsilon _{\mathbf{p}+\mathbf{q}/2}+\epsilon _{\mathbf{p}-
\mathbf{q}/2}}.
\end{equation}
with
\begin{equation}
\label{Dispersion}
\epsilon _{\mathbf{q}}=-2t[\cos(q_x a)+2\cos(q_x a/2)\cos(q_y \sqrt{3} a/2)]
\end{equation}
the non-interacting dispersion relation and $\gamma({\mathbf q})=-\epsilon _{\mathbf{q}}/2t$. In the above formula $\nu$ is the volume of the Brillouin zone (BZ), $f^0({\bf q})=1/[1+\exp((\epsilon_q-\mu_0)/T)]$ is the Fermi function and $\mu_0$ is the non-interacting chemical potential. The pair correlation functions are related to the static structure factors by
\begin{equation}
\label{g_cc}
g_{cc}({\bf r}_{\bf i})=1+\frac{1}{n}\int_{BZ} \frac{d{\bf q}}{\nu}[S_{cc}({\bf q})-1]\exp(i{\bf q}\cdot{\bf r}_{\bf i}),
\end{equation}
\begin{equation}
\label{g_ss}
g_{ss}({\bf r}_{\bf i})=\frac{1}{n}\int_{BZ} \frac{d{\bf q}}{\nu}[S_{ss}({\bf q})-1]\exp(i{\bf q}\cdot{\bf r}_{\bf i}),
\end{equation}
where $S_{cc,ss}({\bf q})=S_{\sigma\sigma}({\bf q})\pm S_{\sigma\tilde{\sigma}}({\bf q})$ are the charge and spin component of the static structure factor. The spin resolved static structure factor is defined by $S_{\sigma\sigma'}({\bf q})=\left\langle n_\sigma({\bf q})n_{\sigma'}({\bf q})\right\rangle/n$ and $n_{\sigma}({\bf q})$ is the Fourier transform of $n_{{\bf i}\sigma}$. The quantities $g_{cc}(a)$ and $g_{ss}(a)$ entering the vertices Eq.(\ref{vertex}) are simply the pair correlation functions at the first-neighbor distance.

Self-consistency is established by connecting the static structure factors to the response functions by the fluctuation-dissipation theorem
\begin{equation}
\label{fdt}
S_{cc,ss}({\bf q})=\frac{T}{n}\sum_{\omega_n}\chi_{cc,ss}({\bf q},\omega_n),
\end{equation}
where $\omega_n=2n\pi T$ are Bosonic Matsubara frequency. Substituting the expression for the susceptibilities Eqs.(\ref{chi_cc} and \ref{chi_ss}) and the corresponding vertices Eq.(\ref{vertex}) on the right-hand side, one can use the result to obtain the pair correlation functions $g_{ss}$ and $g_{cc}$ entering the vertices using their relations Eqs.(\ref{g_cc} and \ref{g_ss}) to the structure factors. Assuming that the functional derivatives of the pair correlation functions are known, as discussed below, we need only three equations to determine the pair correlation functions entering the vertices. This is because the Pauli principle imposes that $g_{\sigma,\sigma}(0)=0$. The equation that is dropped out is that for $g_{cc}(0)=0$. This procedure and its impact on the Pauli principle is discussed in detail in Ref.~\onlinecite{Bahman2}.

Functional derivatives of the pair correlation functions with respect to density and magnetization enter the spin and charge vertices. The functional derivatives are obtained from the following equations:
\begin{align}
\label{fd}
&\frac{\delta g_{s}(1,2)}{\delta n(1)}=[1-g_{cc}(1,2)],\notag\\
&\frac{\delta g_{\uparrow\downarrow}(1,1)}{\delta n(1)}=2[1-g_{\uparrow\downarrow}(1,1)],\\
&\frac{\delta g_{s}(1,2)}{\delta m(1)}=[1-g_{cc}(1,2)].\notag
\end{align}
These equations are strictly valid only when particle-hole symmetry is satisfied. On the square lattice, it has been checked by comparisons with QMC calculations that the results are satisfactory even in the absence of the full particle-hole symmetry~\cite{Bahman2}. Apparently, particle-hole symmetry due to linearization of the dispersion relations near the Fermi surface suffices. We will make this assumption for the triangular lattice where strict particle-hole symmetry is not satisfied. This is justified a posteriori by our results. Those that can be checked against variational QMC, for example, are in excellent agreement.

Finally, the self-energy needed to address the pseudogap problem is obtained following Ref.~\onlinecite{Bahman1}.
\begin{widetext}
\begin{align}
\label{Sigmak}
\Sigma _{\sigma }(\mathbf{k},\omega _{n})\approx  (Un_{\tilde{\sigma}}+6Vn)
&+\frac{T}{4}\sum_{\omega _{n^{\prime}}}\int_{BZ}\frac{d\mathbf{q}}{\nu}\{UU_{ss}(\mathbf{q})\chi _{ss}(\mathbf{%
q},\omega _{n^{\prime }})\notag\\&+U_{cc}(\mathbf{q})[U+4V\gamma(\mathbf{q})]\chi _{cc}(\mathbf{q},
\omega _{n^{\prime }})\}G_{0}(\mathbf{k+q},\omega _{n}+\omega_{n^{\prime }}).
\end{align}%
\end{widetext}
where $\omega_n=(2n+1)\pi T$ is the Fermionic Matsubara frequency and $\omega_n'$ is the Bosonic one. We can also find the spectral function
$A({\mathbf q},\omega)=-\Im G({\mathbf q},\omega)/2\pi$. The above formula does not assume a Migdal theorem since one of the vertices is renormalized. However, it takes into account only the longitudinal fluctuations. Transverse fluctuations could be accounted for following a generalization of the steps in Ref.~\onlinecite{Moukouri:2000}. Since the pseudogap appears only when fluctuations are large, the longitudinal case suffices to establish the qualitative results.

Finally, the interacting chemical potential is obtained from
\begin{equation}
n=T\sum_{\omega _{n}}\int_{BZ}\frac{d\mathbf{q}}{\nu} A(\mathbf{q},\omega_n).
\end{equation}

\section{Response functions}

The ``phase diagrams'' that we present in the section that follows this one, are determined from the growth of the the spin and charge response functions as temperature decreases. When the interaction is local, $U$, the wave vector of the instability is determined entirely from the non-interacting susceptibility, in other words from nesting properties of the Fermi surface. The introduction of near-neighbor repulsion $V$ changes this since it introduces a wave vector dependence to the vertices. In order to disentangle the various effects, we present the non-interacting susceptibilities in the first subsection and the results with interactions in the second subsection. These numerical results are obtained from Eqs.~(\ref{chi_ss}-\ref{fd}).

\subsection{Non-interacting susceptibility}

\begin{figure}[tbp]
\begin{center}
\includegraphics[scale=0.4]{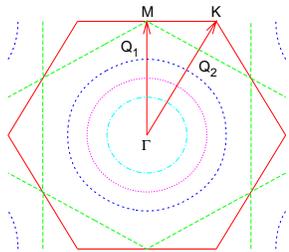}
\end{center}
\caption{(Color online) The non-interacting Fermi surfaces at fillings, starting from the center, $n=0.5,\;1,\;1.25\;{\rm and}\; 1.5$.}
\label{Fig_1}
\end{figure}

\begin{figure}[tbp]
\begin{center}
\includegraphics[scale=0.2]{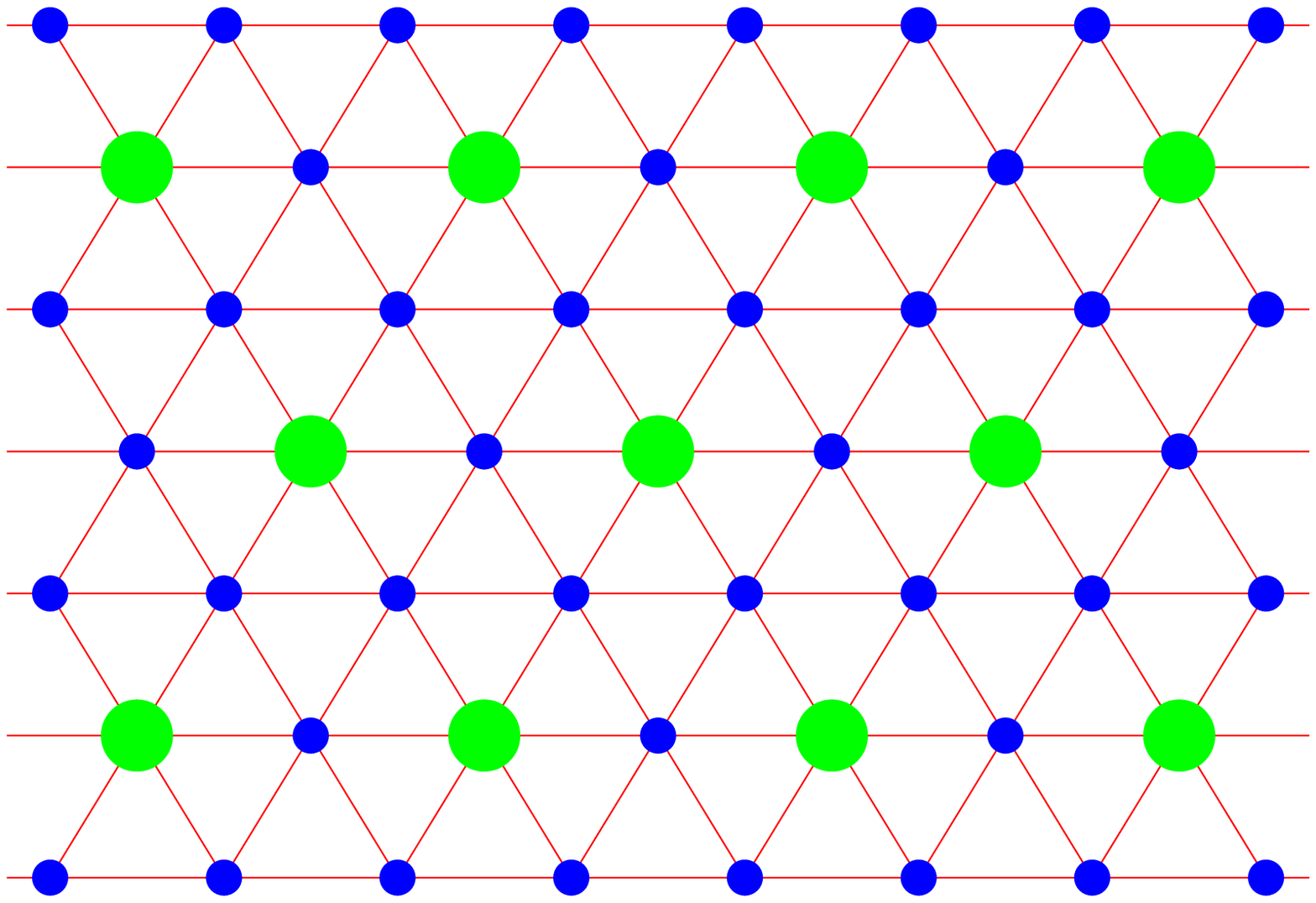}
\includegraphics[scale=0.2]{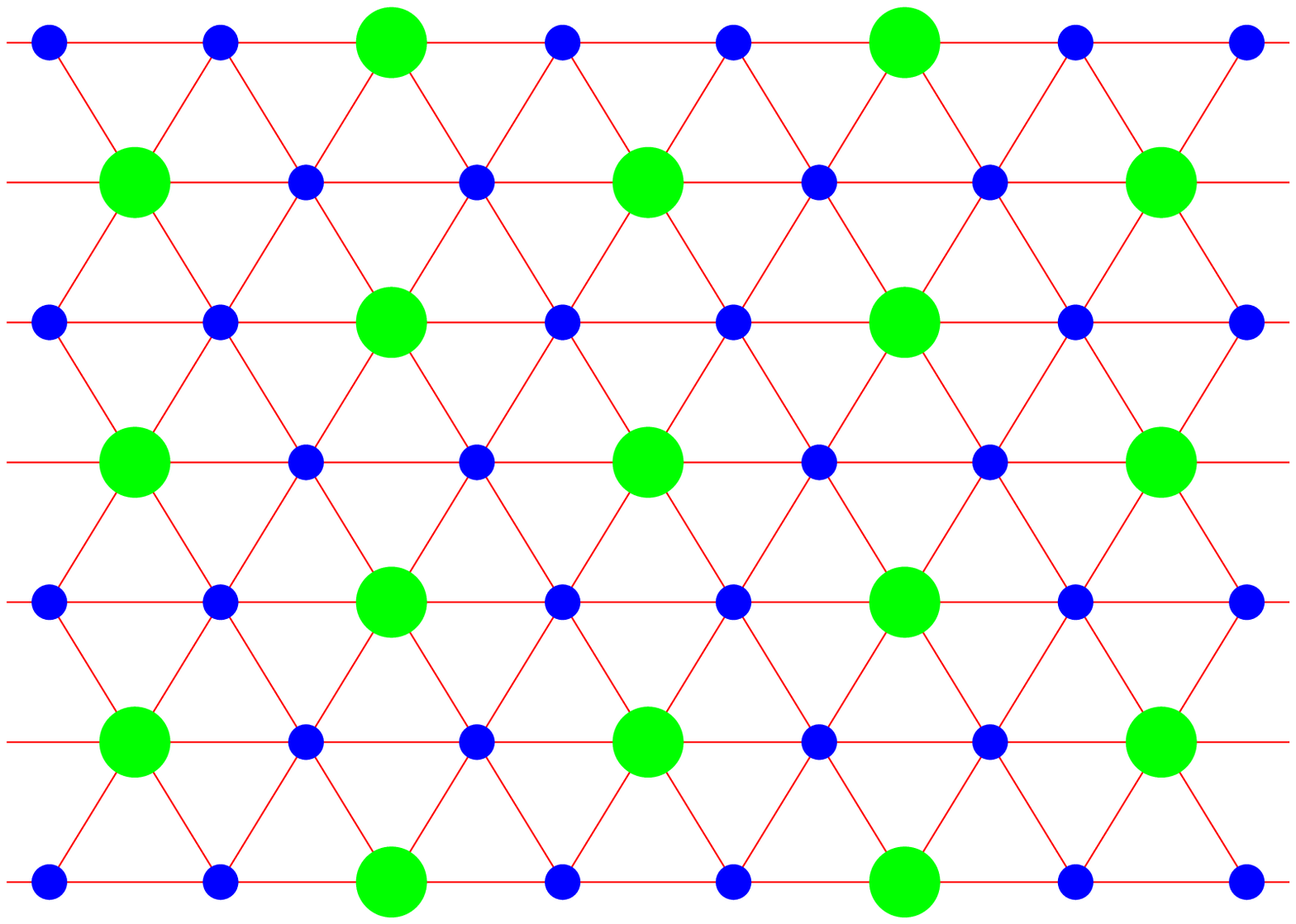}
\end{center}
\caption{(Color online) Real-space structure for two types of order: CDW1 on the left panel and CDW2 on the right panel are related respectively to wave-vectors $Q_1$ and $Q_2$ in Fig.~\ref{Fig_1}.}
\label{Fig_2}
\end{figure}

The non-interacting susceptibility Eq.~\ref{Chi_0} is determined mostly by the shape of the non-interacting Fermi surface that is in turn determined by the dispersion relation given in Eq.~(\ref{Dispersion}). In Fig.~\ref{Fig_1} we present the Fermi surface for increasing values of the density, $n=0.5,\;1,\;1.25\;{\rm and}\;
1.5$ respectively. The first Brillouin zone is plotted as a solid line. It is important to notice that the Fermi surface corresponding to $n=1.5$ (long dash) touches the first Brillouin zone boundary and has long parallel segments that lead to near nesting. We will show in more details that this causes a strong peak in the non-interacting response function. We also draw two wave-vectors that are often found for the most important charge or spin density waves in the system. At these wave-vectors, the charge or spin response functions often have a strong peak. The real-space modulations corresponding to these the wave-vectors are depicted in Fig.~\ref{Fig_2}.

\begin{figure}[tbp]
\begin{center}
\includegraphics[scale=0.4]{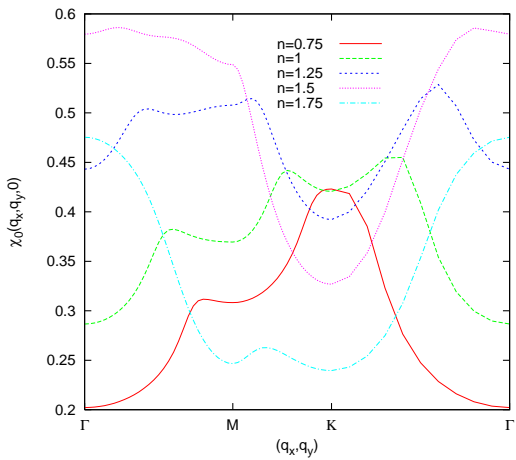}
\end{center}
\caption{(Color online) The free response function for different at $n=0.75,\;1,\;1.25\;1.5\;{\rm and}\; 1.75$ for $T=0.2$}
\label{Fig_3}
\end{figure}

The non-interacting response function Eq.~\ref{Chi_0} is drawn in Fig.~\ref{Fig_3} for different values of densities $n=0.75,\;1,\;1.25\;1.5\;{\rm and}\; 1.75$ at $T=0.2$. The largest response is for $n=1.5$. While one might have expected that parallel segments of the Fermi surface would have lead to a peak at a single dominant wave vector, it seems that the frustration imposes a less pronounced maximum. However, the height of the maximum at that density increases rapidly with decreasing temperature. The position of all the peaks changes only slightly with temperature. The height of the peaks for the smaller values of density does not change drastically with decreasing temperature. That fact in addition to quantum renormalization of the interactions are the main reasons for the absence of any instability at low density up to intermediate coupling.

There is a deep minimum near the K point at higher values of the density. This is the main reason for absence of commensurate spin density wave (SDW). The non-interacting response function has a peak at the commensurate wave-vector $Q_2$ (K point) at lower value of the density but, as we just mentioned, this peak does not grow
enough to produce any sort of order including SDW up to intermediate coupling. That is not the case in the strong coupling limit but that is
out of reach of our approach in the density regime where Mott physics is dominant.

It is quite remarkable that the free response function shows a strong peak at the origin for $n=1.75$, a signature for ferromagnetism at nearby densities. It seems that, as we will see, frustration on the triangular lattice favors ferromagnetism at intermediate coupling, contrary to the square lattice case.

\subsection{Interacting response functions}

\begin{figure}[tbp]
\begin{center}
\includegraphics[scale=0.4]{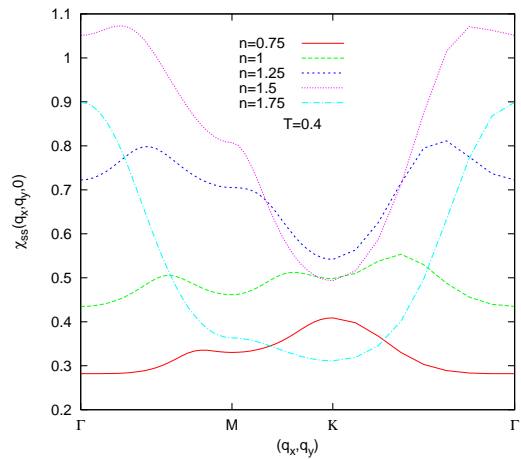}
\end{center}
\caption{(Color online) The spin response function at fixed $U=4$, $V=1.5$, $T=0.4$ and different value of the densities $n=0.75,\;1,\;1.25,\;1.5\;{\rm and}\;1.75$.}
\label{Fig_7}
\end{figure}

\begin{figure}[tbp]
\begin{center}
\includegraphics[scale=0.4]{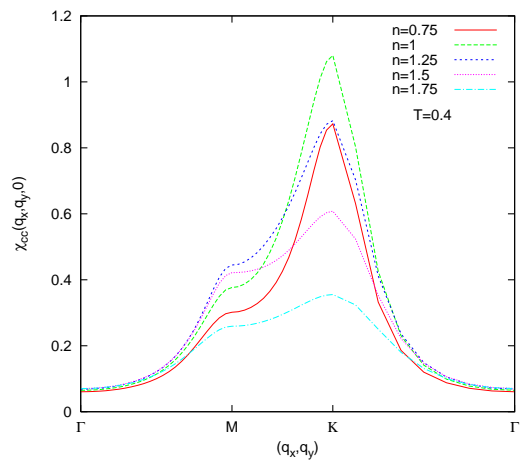}
\end{center}
\caption{(Color online) The charge response function at $U=4$,  $V=1.5$ and $T=0.4$ for different value of the densities.}
\label{Fig_5}
\end{figure}

In the presence of both types of interactions $U$ and $V$, the response functions are strongly modified. Consider typical values of the interaction, $U=4$ and $V=1.5$. Using the same color code as in the figure for the non-interacting case, we show in Fig.~\ref{Fig_7} the spin, Eq.(\ref{chi_ss}), and in Fig.~\ref{Fig_5} the charge, Eq.(\ref{chi_cc}), response functions at $T=0.4$ for the same densities as in the non-interacting case Fig.~\ref{Fig_3}.

In the ordinary random phase approximation, the spin response function in influenced only by the interaction $U$ and the maxima are at the same location as the in non-interacting case. In the present approach however, the nearest-neighbor interaction $V$ also influences the spin response, introducing a wave vector dependent vertex. Hence, some of the maxima of the spin response function in Fig.~\ref{Fig_7} are not at the same wave vector as in the non-interacting case. Nevertheless, the differences are much smaller than for the charge response function appearing in Fig.~\ref{Fig_5}. In the latter case, the position of the maximum is near point K (wave vector $Q_2$) for all densities, in other words the charge response is dominated by the wave vector dependent vertex introduced by $V$.

One can summarize the results for the location of the maximum spin response function as follows. In the range $n\leq 1.5$ the tendency is towards an incommensurate spin-density wave (ISDW) while the response is maximum near zero wave vector (ferromagnetism) just above this density. Generally the height of the peaks increases when $V$ is reduced, hence nearest-neighbor repulsion does not favor spin order. For $n=0.75$, the maximum is near point K corresponding to the same lattice structure as the CDW2 depicted in Fig.~\ref{Fig_2} except that one should replace the big or small points with up and down spins. This resembles the spin structure that would arise with ferrimagnetic order.

For the charge response in Fig.~\ref{Fig_5}, tendency towards CDW2 order (K point) is robust for these values of $U$ and $V$. The tendency is strongest at low values of the density because of a weaker effect of the frustration that leads to a dip in the non-interacting response function in Fig.~\ref{Fig_3}. The effect of frustration is very strong for densities very close to $n=1.5$. But nevertheless the height of the peak for densities around $n=1.5$ grows dramatically with decreasing temperature.

\begin{figure}[tbp]
\begin{center}
\includegraphics[scale=0.4]{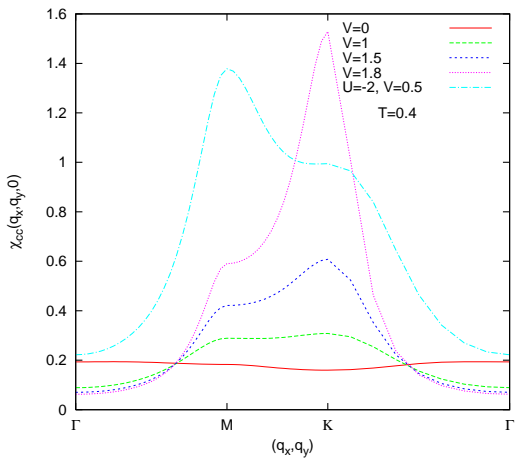}
\end{center}
\caption{(Color online) The charge response function at $n=1.5$ and $T=0.4$ for different value of $U$ and $V$. When not specified, $U$ takes the value $U=4$.}
\label{Fig_4}
\end{figure}

\begin{figure}[tbp]
\begin{center}
\includegraphics[scale=0.4]{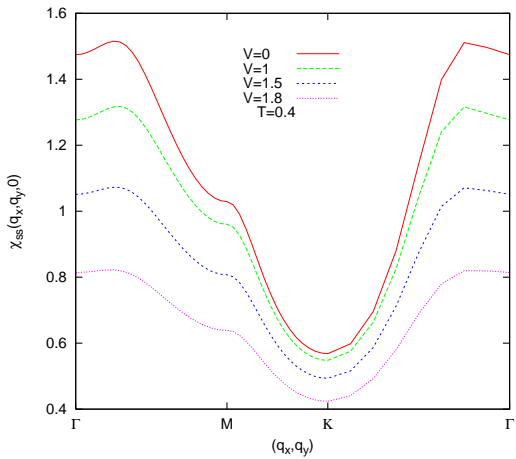}
\end{center}
\caption{(Color online) The spin response function at fixed $n=1.5$, $U=4$, $T=0.4$ and different values of $V=0,\;1,\;1.5\;{\rm and}\;1.8$.}
\label{Fig_6}
\end{figure}

We verify the dominant effects of $U$ and $V$ discussed above, this time by fixing the filling at $n=1.5$ and changing the interactions. Fig.~\ref{Fig_4} and Fig.~\ref{Fig_6} show, respectively, the effect on the charge and spin response functions at $T=0.4$.

For the charge response function in Fig.~\ref{Fig_4}, we imply $U=4$ for those curves where the value of $U$ is not written. A simple
comparison of the charge response function with the corresponding non-interacting susceptibility for $n=1.5$ in Fig.~\ref{Fig_3} shows the importance of the $V$ term in the vertex
Eq.~(\ref{vertex}). The non-interacting susceptibility has a deep minimum on the Brillouin zone boundary. The charge response function on the other hand shows two different maxima at wave-vectors $Q_1$ and $Q_2$. The CDW modulation related to these wave vectors are illustrated in Fig.~\ref{Fig_2}. We will see that these instabilities occur over a wide area in the $U-V$ plane. In fact apart from the phase separation instability ($\mathbf{q=0}$), which occurs for negative $V$, they are the only charge instabilities that can be found at this density. The situation changes as we change the density and one can expect to find incommensurate CDW instabilities in another region of the $U-V$ plane.

Moving on to the spin susceptibility, Fig.~\ref{Fig_6} shows that the presence of the $V$ term suppresses the spin response function more and more as $V$ increases, concomitant with the increase in the charge response function. In principle,
one cannot find a strong maximum in both the spin and charge response functions. This is true in all one band homogeneous paramagnetic systems as dictated by Eqs.~(\ref{g_cc}), (\ref{g_ss}) and $g_{\sigma\sigma}(0)=0$ (Pauli sum-rule). Indeed, the Pauli sum-rule (obtained from the Pauli principle $<n^2_{\sigma}>=<n_{\sigma}>$) connects $g_{cc}(0)$ to $g_{ss}(0)$ in such a way that an increase in one forces a decrease in the other~\cite{AM1}.

\section{Crossover diagrams}

In mean-field theory, one normally finds finite temperature phase transitions, in contradiction with the Mermin-Wagner theorem. In ETPSC, we obtain instead at a temperature $T_X$ below which the correlation length begins to grow exponentially, diverging only at zero temperature. $T_X$ is lower than the mean-field transition temperature because of the quantum Kanamori-Br\"{u}ckner renormalization of the vertices. In this low temperature regime, the characteristic frequency of the growing fluctuations becomes less than temperature in dimensionless units. This is the renormalized-classical regime. Either the spin or the charge correlations grow exponentially at some characteristic wave vector that suggests which long-range order will likely be stabilized at zero temperature. Since our approach is not valid deep in the renormalized classical regime, one cannot be sure that the zero-temperature phase will be precisely that suggested by the behavior at $T_X$.

The value of $T_X$ depends on density, $U$ and $V$. We use $\chi({\mathbf q}_x,{\mathbf q}_y,0)/\chi_0({\mathbf q}_x,{\mathbf q}_y,0)={\rm const}$ to estimate $T_X$. For the sake of computational efficiency, we chose the constant to be $10$ and checked that the general features do not change by choosing a larger value. This occurs because the exponential growth of the correlations is rather sudden. A detailed discussion of this issue can be found in previous publications~\cite{Bahman1,Bahman2}.

\begin{figure}[tbp]
\begin{center}
\resizebox{7 cm}{6.5 cm}{\includegraphics[bb=0 60 600 620]{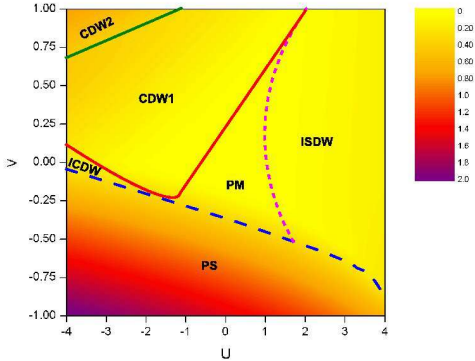}}
\end{center}
\caption{(Color online) Value of the crossover temperature $T_X$ to the renormalized classical regime as a function of $U$ and $V$ at filling $n=1.5$. The wave vector and spin or charge character of the growing correlations is indicated by initials: Incommensurate spin density waves (ISDW), phase separation, or $\mathbf{q}=0$ charge instability (PS), and incommensurate charge density waves (ISDW). The wave vectors of the two special charge density waves CDW1 and CDW2 (both $\sqrt(3) \times \sqrt(3)$) are shown in Figs.~\ref{Fig_1} and \ref{Fig_2}. The color scale (grey scale) appears on the right of the plot. Regions where either $U$ or $V$ are negative are shown for illustrative purposes only.}
\label{Fig_8}
\end{figure}

\begin{figure}[tbp]
\begin{center}
\resizebox{6 cm}{5 cm}{\includegraphics[bb=0 120 600 620]{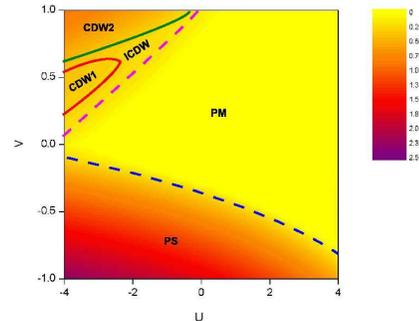}}
\end{center}
\caption{(Color online) Crossover temperature $T_X$ to the renormalized classical regime as a function of $U$ and $V$ at filling $n=4/3$. Other symbols are defined in the caption of Fig.~\ref{Fig_8}. Regions where either $U$ or $V$ are negative are shown for illustrative purposes only.}
\label{Fig_9}
\end{figure}

\begin{figure}[tbp]
\begin{center}
\resizebox{6 cm}{5 cm}{\includegraphics[bb=0 120 600 620]{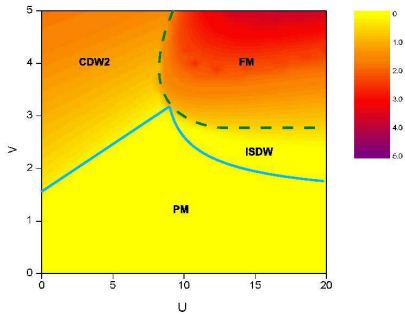}}
\end{center}
\caption{(Color online) Crossover temperature $T_X$ to the renormalized classical regime as a function of $U$ and $V$ at filling $n=2/3$. Other symbols are defined in the caption of Fig.~\ref{Fig_8}. The only new acronym, FM stands for ferromagnetic. Results for the regions where $U$ and $4V$ are larger than about half the bandwidth are presented only for illustrative purposes.}
\label{Fig_11}
\end{figure}

\begin{figure}[tbp]
\begin{center}
\resizebox{6 cm}{ 5 cm}{\includegraphics[bb=0 120 600 620]{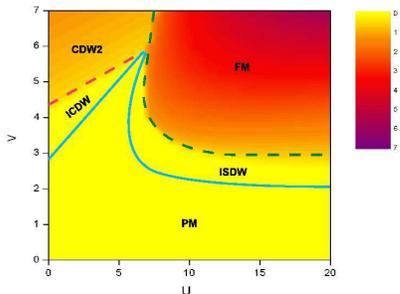}}
\end{center}
\caption{(Color online) Crossover temperature $T_X$ to the renormalized classical regime as a function of $U$ and $V$ at filling $n=0.5$. Other symbols are defined in the caption of Figs.~\ref{Fig_8} and \ref{Fig_11}. Results for the regions where $U$ and $4V$ are larger than about half the bandwidth are presented only for illustrative purposes.}
\label{Fig_10}
\end{figure}

\begin{figure}[tbp]
\begin{center}
\resizebox{7 cm}{6 cm}{\includegraphics[bb=0 120 600 620]{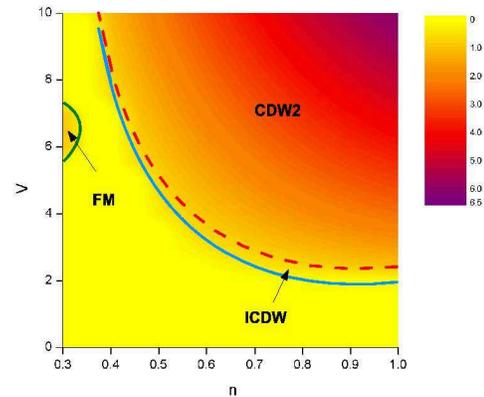}}
\end{center}
\caption{(Color online) Crossover temperature $T_X$ to the renormalized classical regime as a function of $V$ and $n$ at $U=5$. The yellow part of the figure represents a paramagnetic, or Fermi liquid, regime. Figs.~\ref{Fig_8} and ~\ref{Fig_11} for the meaning of symbols.}
\label{Fig_12}
\end{figure}

\begin{figure}[tbp]
\begin{center}
\resizebox{5 cm}{6 cm}{\includegraphics[bb=200 120 550 600]{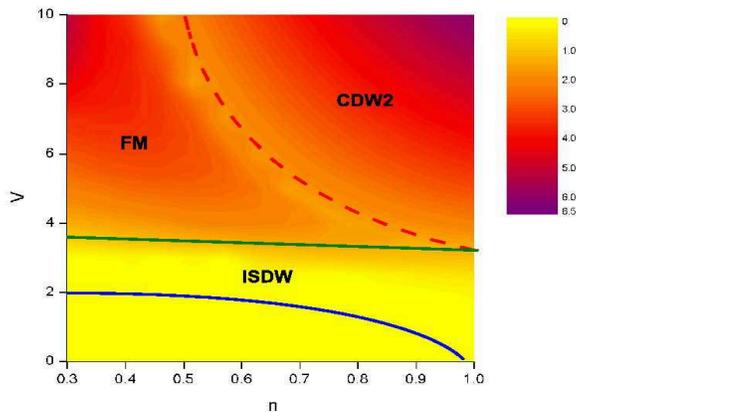}}
\end{center}
\caption{(Color online) Same as Figs.~\ref{Fig_12} but for $U=10$.}
\label{Fig_13}
\end{figure}

We present our results for $T_X$ in Figs.~\ref{Fig_8} to \ref{Fig_13} as color (grey scale) plots in various planes of parameter space. There are four plots that present the $U-V$ dependence of $T_X$ at four densities, then two plots for the $V-n$ dependence at fixed $U$. We indicate by lines of various colors and types the boundaries between regions where there is either a change in the wave vector of the growing correlations, or a change in the type of correlation, spin or charge. When we indicate a paramagnetic (Fermi liquid) region (PM), we mean that correlations did not grow, in either the spin or charge sectors, at temperatures as low as $T=0.01$.

Fig.~\ref{Fig_8} displays $T_X$ at $n=1.5$ as a function of $U$ and $V$ for both positive and negative values. At negative values of either $V$ or $U$, superconducting correlations will be competing. Since superconductivity has not been taken into account here, the results in all quadrants, except the first one, should be taken as just indicative of what may happen in the spin or charge sectors. When $V$ is negative, unless $U$ is large, there is a strong tendency to phase separation (PS), i.e. the static charge response function starts, at $T_X$, to grow exponentially for wave vector $\mathbf{q}=0$. At positive $U$ and $V$, incommensurate spin density waves (ISDW) are dominant, but $U$ and $V$ must be large enough, as expected from the absence of perfect nesting. At small $U$ and $V$, the system remains paramagnetic (PM). Charge density waves appear at positive $U$ and $V$ only if $V$ is relatively large. Recall however that the effect of $V$ is amplified by the presence of several neighbors. The charge instability in this parameter range is of the CDW1 type illustrated in Fig.~\ref{Fig_2}. Charge instabilities are further amplified at positive $V$ only if $U$ is allowed to become negative. The $\sqrt(3) \times \sqrt(3)$ CDW2 pattern is allowed only in extreme conditions of large positive $V$ and large negative $U$. This is not suprising since one can see from the non-interacting Fermi surface in Fig.~\ref{Fig_1} that the corresponding wave vector $Q2$ is not particularly favored by nesting. The CDW2 phase is really governed by properties of the vertex $V$, not so much by the non-interacting Fermi surface.

When density is decreased to $n=4/3$, the non-interacting Fermi surface becomes almost circular so at positive $U$ and $V$ the tendency to order is strongly suppressed, as can be seen from Fig.~\ref{Fig_9}. Compared with the previous figure, the CDW2 vertex related instability is more robust while the CDW1 and ICDW instabilities occur in smaller regions, the ICDW existing over a larger region this time than CDW1. At larger fillings, $n=5/3$, where again the Fermi surface becomes almost circular, similar features are observed.

As the filling decreases, the Fermi surface becomes more and more circular. Restricting ourselves to positive $U$ and $V$, and staying at weak to intermediate coupling where our theory is valid, nothing interesting occurs. The system remains paramagnetic down to $T=0.01$. We thus also present, in Fig.~\ref{Fig_11} and Fig.~\ref{Fig_10}, results at large $U$ and $V$ where our theory is not strictly controlled. We feel these results are nevertheless interesting for two reasons. First, some of our ``phase boundaries'' compare favorably with results obtained from other methods. Second, the renormalized classical regime occurs at such high temperature that $U/T$ and $V/T$ may begin to control the approach.

Fig.~\ref{Fig_11} and Fig.~\ref{Fig_10} thus show the crossover diagram for, respectively, $n=2/3$ and $0.5$ over a wide range of positive $U$ and $V$. The CDW2 region now appears at positive $U$ and $V$, contrary to the results in the previous figures, as long as the stabilizing interaction $V$ is large enough. The boundary that separates CDW2 from PM in Fig.~\ref{Fig_11} is very close to QMC results~\cite{Watanabe}, which gives us confidence in the validity of the results. The ferromagnetic phase is dominant when both $U$ and $V$ are large. The ICDW phase does not appear at filling $n=2/3$ (Fig.~\ref{Fig_10}). The CDW2 phase is influenced to some extent not only by the vertex, but also by commensurability, as can be seen from the fact that it is more important at $n=2/3$ than at $n=0.5$. There are competing tendencies for the CDW2 phase: 1) The density $n=2/3$ is more favorable for the commensurate CDW2 as reflected in the free response function, 2) The effect of off-site interaction is less important at lower value of the density. Based on these observations, one might surmise the presence of an optimal density where CDW2 appears over a larger area of $U-V$ space.

To explore in more details the density dependence, we present results as a function of $n$ and $V$ at fixed $U$. This was studied in particular by Motrunich and Lee~\cite{Motrunich1,Motrunich2}. They calculated the phase diagram with different methods: a) renormalized mean field theory and variational quantum Monte Carlo with a trial wave function~\cite{Motrunich1} and b) slave boson mean field theory~\cite{Motrunich2}. They suggest that the effect of the $V$ term is taken into the account more accurately in the first method than in the second one. In the first paper, they found that the CDW2 phase can be reached at smaller $V$ at the densities $n=1/3$ and $2/3$ than at other densities. This is not the case in the second paper where these densities play no special role. Since their calculations are at $U\rightarrow\infty$, this suggests that the general features of the phase diagram can be understood physically at low and high density: One needs a large $V$ to stabilize the CDW at low value of the density. Since at large value of the density, the hole plays the same role as the electron at low density, we expect at large $V$ to find the CDW as well. This argument is correct if other phases do not suppress the CDW.

We present the results of calculations at finite values of $U$ in Figs.~\ref{Fig_12} and \ref{Fig_13}. In Fig.~\ref{Fig_12}, the results in the $V-n$ plane are for $U=5$. It can easily be seen that CDW2 appears when both $V$ and $n$ are large. A ICDW region separates CDW2 from the paramagnetic, or Fermi liquid phase. There is a small area on the left of the plane where the ferromagnetic phase is stable. Fig.~\ref{Fig_13} shows the results for $U=10$. CDW2 is still stable in the same region of the $V-n$ plane but the small FM region of the previous figure has now grown and pushed away slightly the CDW2 phase. In other words, at larger $U$, spin fluctuations are playing a more important role, as expected. This is also shown by the appearance of an ISDW regime. It is obvious from Figs.~\ref{Fig_12} and \ref{Fig_13} that the densities $n=1/3$ and $2/3$ do not play any special role, at least at these values of $U$. This is in agreement with the results of Ref.~\onlinecite{Motrunich2}. However, at large $U$, we find more ferromagnetic spin fluctuations, a possibility that was not considered in Ref.~\onlinecite{Motrunich2}.  We also calculated the same phase diagram at higher value of $U$ where we find that the CDW2 region is completely swept away by ferromagnetism.

\section{Fermi surface and pseudogap}

It has already been shown that spin and superconducting thermal fluctuations can~\cite{AM1,AM2,Kyung:2001,Dare:2004}, in two dimensions, open up a pseudogap on the Fermi surface that reflects the wave vector of the fluctuations. The same study can be performed here to check the effect of charge density wave fluctuations. We obtain the self-energy by substituting our results for the susceptibilities in Eq.~(\ref{Sigmak}), from which one can compute the spectral function.

\begin{figure}[tbp]
\centering
\hspace{-3cm}
\includegraphics[scale=0.6]{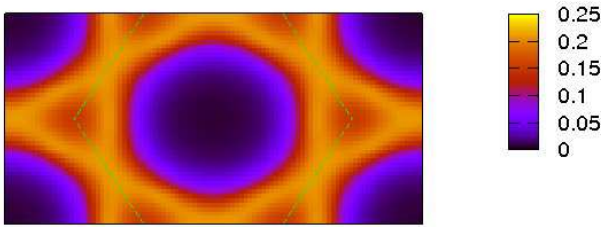}
\caption{(Color online) The spectral function in the Fermi liquid regime as a function of wave-vector at $\omega=o$, $n=1.5$, $U=4$, $V=0$ and $T=0.4$. The dashed Green line is the Brillouin zone.}
\label{Fig_14}
\end{figure}

We begin by the effect of spin fluctuations. In Fig.~\ref{Fig_14}, we plot $A({\mathbf q},\omega=0)$ for $n=1.5$, $U=4$, $V=0$ and $T=0.4$. In the jargon, this is known as a Momentum distribution curve (MDC). The dashed Green line is the Brillouin zone. Following the largest intensity regions, one can recognize the shape of the non-interacting Fermi surface illustrated for $n=1.5$ in Fig.~\ref{Fig_1}. It is clear from this figure that the effect of the on-site interaction at this temperature is just to introduce damping. Since this is a region where ISDW appear at low temperature, the Fermi surface can be destroyed by lowering the temperature or increasing $U$.

\begin{figure}[tbp]
\centering
\resizebox{7 cm}{6 cm}{\includegraphics[bb=80 40 250 220]{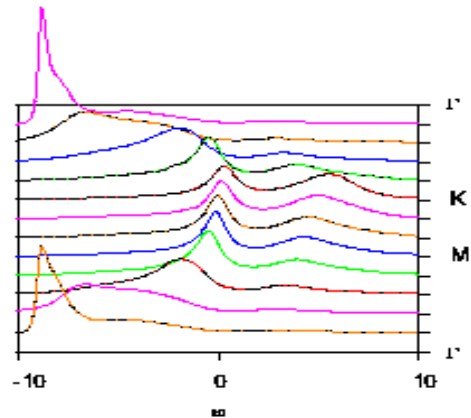}}
\caption{(Color online) The spectral function as a function of $\omega$ at $n=1.5$, $U=4$, $V=0$ and $T=0.4$ for different values of the wave-vector. Symmetry points are defined in Fig.~\ref{Fig_1}.}
\label{Fig_16}
\end{figure}

For better understanding, we plot in Fig.~\ref{Fig_16} the spectral function as a function of frequency at different wave vectors for the same parameters as in the previous figure. These are Energy dispersion curves (EDC). One can clearly observe the quasi-particle dispersion relation, the effect of $U$ appearing as damping. The extra features at higher frequency are precursors of extra bands that would appear in the ordered state. Indeed, the EDC are plotted in a regime where the correlation length associated with incommensurate fluctuations is three to four lattice spacings ($\chi({\mathbf q}_x,{\mathbf q}_y,0)/\chi_0({\mathbf q}_x,{\mathbf q}_y,0)=10$), enough to enter the renormalized classical regime. A simple generalization of an argument presented earlier~\cite{AM2} shows that in that regime, the self-energy can be approximated by
\begin{equation}
\Sigma\left(  \mathbf{k,}\omega\right)  =\sum_{i=1}^{n}\frac{\Delta^{2}%
}{\omega-\varepsilon_{\mathbf{k+q}_{i}}+i\Gamma}%
\end{equation}
where the sum runs over all the equivalent maxima of the susceptibility, six
of them for the present case. Substituting this expression into the general
result for the spectral weight%
\begin{equation}
A\left(  \mathbf{k,}\omega\right)  =\frac{-2\Sigma^{\prime\prime}\left(
\mathbf{k},\omega\right)  }{\left(  \omega-\varepsilon_{\mathbf{k}}%
-\Sigma^{\prime}\left(  \mathbf{k},\omega\right)  \right)  ^{2}+\Sigma
^{\prime\prime}\left(  \mathbf{k},\omega\right)  ^{2}}%
\end{equation}
one can reproduce qualitatively the behavior in Fig.~\ref{Fig_16}. With
larger correlation length (larger $\Delta$), the spectrum clearly splits into
more bands.

\begin{figure}[tbp]
\centering
\includegraphics[scale=0.45]{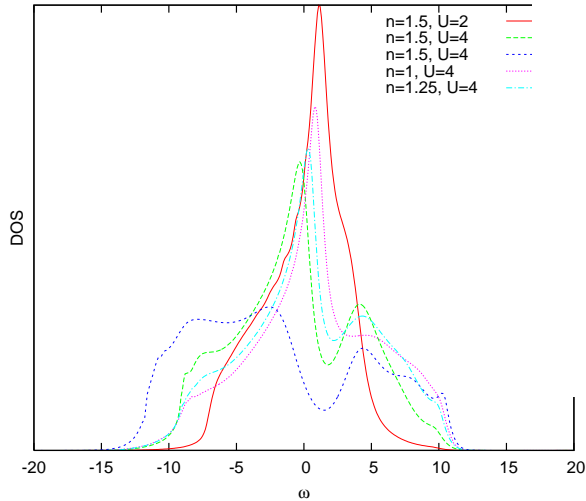}
\caption{(Color online) The density of state for different values of the densities and $U$ for $V=0$, $T=0.4$.}
\label{Fig_18}
\end{figure}

The spin fluctuation induced precursor effects are also illustrated in Fig.~\ref{Fig_18} that displays the density of states for various values of $U$ and for densities near $n=1.5$. The $U=2$ curve is close to the non-interacting result. For stronger interaction, the extra band of states can be understood as a precursor effect by drawing the density of states for a static modulation.

\begin{figure}[tbp]
\centering
\hspace{-3cm}
\includegraphics[scale=0.6]{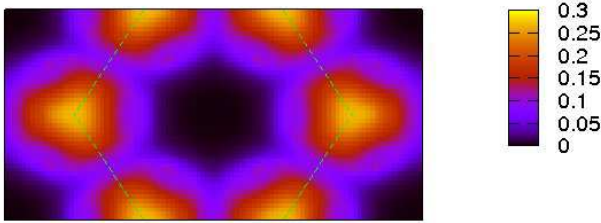}
\caption{(Color online) The spectral function as a function of wave-vector at $\omega=\mu$, $n=1.5$, $U=4$, $V=1.95$ and $T=0.4$}
\label{Fig_15}
\end{figure}

The fact that renormalized classical spin fluctuations can create a pseudogap on the Fermi surface in two dimensions is well documented so we do not show a figure corresponding to this case. We present the case of a Fermi surface pseudogap that originates from charge fluctuations. Increasing $V$ enough that CDW1 fluctuations become important, one can observe a pseudogap, as seen in $A({\mathbf q},\omega=0)$ for $n=1.5$, $U=4$, $V=1.95$ and $T=0.4$ displayed in Fig.~\ref{Fig_15}. The pseudogap opens up at the $M$ point, in other words the spectral function at this point is suppressed. By analogy with the spin-fluctuation induced pseudogap, the symmetry equivalent $M$ points are nearly connected by the symmetry equivalent wave vectors $Q1$ of the CDW1 charge instability.

\begin{figure}[tbp]
\centering
\resizebox{7 cm}{6 cm}{\includegraphics[bb=80 0 250 160]{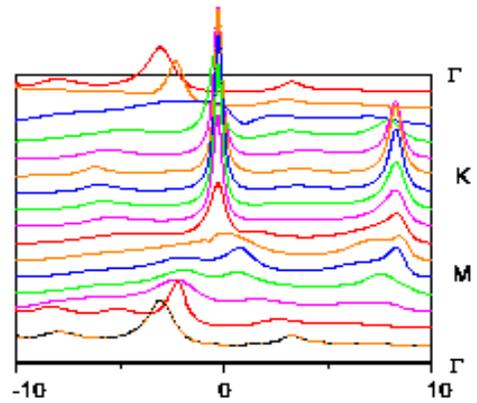}}
\caption{(Color online) The spectral function as a function of  $\omega$ at $n=1.5$, $U=4$, $V=1.95$ and $T=0.4$ for different value of the wave-vector. Symmetry points are defined in Fig.~\ref{Fig_1}.}
\label{Fig_17}
\end{figure}

The EDC corresponding to this situation are illustrated in Fig.~\ref{Fig_17}. It is clear again from this figure that the pseudogap opens up around the $M$ point where the suppression occurs in the MDC of Fig.~\ref{Fig_15}. For other points the peak positions change only slightly.

\section{Discussion and conclusion}

We have used ETPSC to clarify the leading instabilities of the extended Hubbard model on the triangular lattice. 
The method is valid from weak to intermediate coupling. It satisfies the Mermin-Wagner theorem, includes quantum 
(Kanamori-Br\"{u}uckner) renormalization of the vertices and does not assume a Migdal theorem for the self-energy. 
We interpret the entry into the renormalized classical regime (exponential growth of the correlation length) as an 
indication of the phase that acquires long-range order at zero temperature. Since we scan all wave vectors for both 
spin and charge instabilities, our method is not biased towards a restricted set of instabilities as most other studies. 
Superconducting instabilities, however, have not been explored.

The range of possible phases as a function of on-site interaction $U$, nearest-neighbor interaction $V$ and filling $n$ 
is quite rich. Notwithstanding superconducting instabilities, negative values of $U$ and $V$ favor charge instabilities 
either at zero wave vector (phase separation) when $U$ dominates, or at the CDW1 and CDW2 wave vectors when $V$ dominates.
In the physically more relevant regime where $U$ and $V$ are both positive, spin instabilities are favored when $U$ 
dominates and they occur at wave vectors that are essentially determined by the Fermi surface. That is particularly clear 
at large fillings where the shape of the Fermi surface is non-trivial. Larger $V$, on the other hand, favors charge 
instabilities that are generally determined by the vertex itself instead of by details of the Fermi surface. This is particularly clear at small filling where the Fermi surface is essentially circular. That predominance of the vertex is why the CDW2 ($\sqrt(3) \times \sqrt(3)$) phase, for example, exists over a wide range of fillings and is not favored by commensurate fillings. We find that this CDW2 phase can compete with spin instabilities, especially ferromagnetism, when $U$ is large as well. Ferromagnetic fluctuations appear in a range of doping similar to that observed for the cobaltates. That competition with ferromagnetism has not been taken into account in earlier studies~\cite{Motrunich1,Motrunich2}. All the phases, except the Fermi liquid one, appear at large values of the interactions when filling is small, somewhat outside the regime of validity of our approach. Nevertheless, agreement with other approaches suggest that ETPSC extrapolates in a reasonable way towards strong coupling. The disappearance of all phases except the Fermi liquid one at small coupling and small fillings ($n=2/3$) on the triangular lattice is a clear manifestation of the effects of frustration, as can be seen by contrasting with the square lattice case~\cite{Bahman2} ($n=0.75$) where this does not occur.

Finally, we also showed that charge-density wave thermal fluctuations can also induce a pseudogap. A pseudogap associated with CDW is observed~\cite{Kordyuk:2008} experimentally. If its origin is the one discussed in the present paper, it should disappear as temperature rises above that where the charge correlation length becomes of the order of the thermal de Broglie wave length~\cite{AM2}, as observed in the spin fluctuation case in electron-doped cuprates~\cite{Motoyama:2007}.

Further studies should include the competition with superconducting fluctuations as well as more realistic modeling of the specific materials to which one wishes to apply our results. For example, the $n=1$ case is relevant for the layered organics.

\section{Acknowledgments}
Computations were performed on the Elix2 Beowulf cluster in Sherbrooke and on the Ms cluster of the R\'eseau Qu\'eb\'ecois de calcul haute performance (RQCHP). The present work was supported by the Natural Sciences and Engineering Research Council NSERC (Canada), the Fonds qu\'{e}b\'{e}cois de recherche sur la nature et la technologie FQRNT (Qu\'{e}bec), the Canadian Foundation for Innovation CFI
(Canada), the Canadian Institute for Advanced Research CIFAR, and the Tier I
Canada Research Chair Program (A.-M.S.T.).



\end{document}